\documentclass[final,5p,times,twocolumn]{elsarticle}

\usepackage{placeins}
\usepackage{lipsum}
\usepackage{xcolor}
\usepackage[normalem]{ulem}
\usepackage{comment}
\usepackage{tabularx}
\usepackage{graphicx}
\usepackage{dcolumn}
\usepackage{bm}
\usepackage{bbm}

\usepackage{blindtext}
\usepackage{graphics}
\usepackage{verbatim}
\usepackage{amsfonts}
\usepackage{amsmath}
\usepackage{amssymb}
\usepackage{adjustbox}
\usepackage{microtype}
\allowdisplaybreaks
\usepackage{xspace} 
\usepackage{xparse} 
\usepackage{multirow} 
\usepackage{tabstackengine}
\usepackage{algorithm}
\usepackage{algorithmic}

\usepackage{float}
\usepackage{booktabs}

\usepackage[utf8]{inputenc}
\usepackage[T1]{fontenc}
\setstackEOL{\cr}

\newcommand{\Eq}[1]{Eq.~\eqref{#1}}

\newcommand{\Fig}[1]{Fig.~\ref{#1}}

\newcommand{\Alg}[1]{Alg.~\ref{#1}}

\def\md{\mathrm{d}}

\newcommand*{\ndots}{\kern-0.075em.\kern-0.05em.\kern-0.05em.}
\newcommand*{\nidots}{.\kern-0.05em.\kern-0.05em.}
\newcommand*{\ncdots}{\kern-0.15em\cdot\kern-0.2em\cdot\kern-0.2em\cdot\kern-0.15em}  

\NewDocumentCommand{\doubleI}{O{}}{\mathbbm{1}_{#1}}
\NewDocumentCommand{\doubleIb}{O{}}{{\overline{\mathbbm{1}}_{#1}}}
\NewDocumentCommand{\doubleIk}{O{}}{\mathbbm{1}^\ks_{\! #1}}
\NewDocumentCommand{\doubleId}{O{}}{\mathbbm{1}^\ds_{\! #1}}
\NewDocumentCommand{\doubleIp}{O{}}{\mathbbm{1}^\ps_{\! #1}}
\NewDocumentCommand{\doubleV}{O{}}{\mathbbm{V}_{\! #1}}
\NewDocumentCommand{\doubleVk}{O{}}{\mathbbm{V}^\ks_{\! #1}}
\NewDocumentCommand{\doubleVd}{O{}}{\mathbbm{V}^\ds_{\! #1}}
\NewDocumentCommand{\doubleVp}{O{}}{\mathbbm{V}^\ps_{\! #1}}
\NewDocumentCommand{\doublev}{o}{{\mathbbm{v}_{#1}}}
\NewDocumentCommand{\doubleVb}{o}{{\overline{\mathbbm{V}}_{\! #1}}}
\NewDocumentCommand{\doubleVt}{o}{{\widetilde{\mathbbm{V}}_{\! #1}}}
\NewDocumentCommand{\doubleVh}{o}{\widehat{{\mathbbm{V}}_{\! #1}}}
\NewDocumentCommand{\doubleW}{o}{\mathbbm{W}_{\! #1}}
\NewDocumentCommand{\doubleWk}{o}{\mathbbm{W}^\ks_{\! #1}}
\NewDocumentCommand{\doubleWd}{o}{\mathbbm{W}^\ds_{\! #1}}
\NewDocumentCommand{\doubleWb}{o}{{\overline{\mathbbm{W}}_{\! #1}}}
\NewDocumentCommand{\doubleWt}{o}{{\widetilde{\mathbbm{V}}_{\! #1}}}
\NewDocumentCommand{\doubleWh}{o}{{\widehat{\mathbbm{V}}_{\! #1}}}

\newcommand{\Saitama}{Department of Physics, Saitama University, 338-8570, Japan}

\definecolor{darkgreen}{rgb}{0,0.5,0}
\definecolor{purple}{rgb}{0.6,0,0.5}
\definecolor{purple}{rgb}{0.35,0,0.35}
\definecolor{orange}{rgb}{1,0.5,0}
\definecolor{darkred}{rgb}{.7,0,0}
\definecolor{darkblue}{rgb}{0,0,.6}
\definecolor{grey}{rgb}{.6,.6,.6}
\definecolor{dimgreen}{rgb}{0.2,0.7,0.2}
\newcommand{\Lsub}{{\mathcal{L}_\text{sub}}}

\usepackage{multirow}
\usepackage{physics}
\usepackage{hyperref}
\hypersetup{colorlinks=true,breaklinks,linkcolor=blue,urlcolor=blue,citecolor=blue}

\begin{document}

\begin{frontmatter}

\title{Parallelized contraction of tensor trains or matrix product operators}

\author[LMU]{Simone Foderà}

\affiliation[LMU]{organization={Arnold Sommerfeld Center for Theoretical Physics, Center for NanoScience, and Munich Center for Quantum Science and Technology, Ludwig-Maximilians-Universit\"at M\"unchen, 80333 Munich, Germany}}

\author[LMU,CCQ]{Marc K. Ritter}

\affiliation[CCQ]{organization={Center for Computational Quantum Physics, Flatiron Institute, 162 5th Avenue, New York, NY 10010, USA}}

\author[Saitama]{Hiroshi Shinaoka}
\affiliation[Saitama]{organization={\Saitama}}

\author[LMU]{Jan von Delft}

\date{22.06.2026}

\begin{abstract}
Tensor Trains (TT), also known as Matrix Product States (MPS) and Matrix Product Operators (MPO), provide a compact and structured representation for high-dimensional data and operators. One of the most expensive manipulations involving tensor trains is the contraction of two MPOs. A popular and accurate method for mitigating this cost is the fit algorithm. 
However, it is still comparatively costly since it involves 2-site updates. Moreover, the parallelization of the fit algorithm when used for MPO-MPO contractions has received comparatively little attention. 
In this work, we present two strategies for accelerating the fit algorithm, usable in combination: 
(1) We use MPI-based distributed-memory parallelization tailored for MPO-MPO contractions, employing one of two MPO gauge choices: 
(1a) the inverse canonical gauge, which yields near-ideal parallelization speedup across all problem sizes;  and (1b) the site-canonical gauge, which avoids the inversion of singular values but requires extra computations to ensure global consistency, 
thus yielding excellent parallelization speedup only for large problems requiring several sweeps before convergence.
(2) We use randomized projections to reduce the cost of local updates from 2-site to 1-site costs while retaining 2-site accuracy, and to speed up contractions of environment tensors with MPO tensors.

\end{abstract}


\end{frontmatter}

\section{Introduction}\label{sec:introduction}

Tensor trains (TT) are a powerful numerical tool for representing and manipulating high-dimensional or high-resolution functions in a compact and structured way \cite{TT, ODlogN, Ritter2024}. Originally introduced in quantum physics, they are commonly referred to as matrix product states (MPS) when representing physical states or mathematical vectors, and as matrix product operators (MPO) when representing linear operators \cite{MPS}.
We will use the terms TT and MPO interchangeably.

In many applications, it is essential to concatenate operators. When these operators admit efficient MPO representations, this task can be accelerated, sometimes even exponentially, by contracting the MPOs. Yet, in practice, MPO-MPO contraction often remains the computational bottleneck for TT applications. This stems from the fact that while most operations on TTs with bond dimension \(\chi\) scale as $\mathcal{O}(\chi^2)$ or $\mathcal{O}(\chi^3)$ in runtime and require $\mathcal{O}(\chi^2)$ memory space, MPO-MPO contraction scales as $\mathcal{O}(\chi^4)$ in runtime and as $\mathcal{O}(\chi^3)$ in memory consumption \cite{2sFit,XTRG, nunez_fernandez_learning_2025}.
The two algorithms used most commonly are the `zip-up' algorithm introduced by Stoudenmire et al.\ \cite{2sFit} and the `fit' algorithms introduced in different settings by Verstraete and Cirac \cite{1sFit}, and by Chen et al.\ \cite{XTRG}. Recently, new algorithms such as the Successive Randomized Contraction (SRC) \cite{SRC} for MPO-MPS contraction with runtime that scales as $\mathcal{O}(\chi^4)$, and the Recursive Sketched Interpolation (RSI) \cite{RSI} and Alternating Cross Interpolation (ACI) \cite{ACI} for elementwise multiplication with runtime that scale as $\mathcal{O}(\chi^3)$ have been developed, but little attention has been given to MPO-MPO contraction. 

In this work we focus on accelerating the fit algorithm through parallelization. The basic idea,
as simple as it is generic \cite{PDMRG}, is to \textit{partition} the TTs to be contracted into several sub-TTs, \textit{parallelize} the sub-TT contractions, and finally \textit{reconnect} contracted sub-TT neighbors. This approach can be straightforwardly combined with any other strategy that accelerates TT-TT contractions, now applied to sub-TT contractions. Here, we illustrate this by using randomized projections to accelerate the local updates needed for MPO-MPO contractions. 

More specifically, we combine four ingredients for
accelerating the fit algorithm \cite{2sFit,1sFit} and improving its numerical stability:
(i) We partition the TT into several sub-TTs and contract these in parallel \cite{PDMRG}.  (ii) We show how contracted sub-TTs can be reconnected without inverting small singular values (as done in \cite{PDMRG}), thus improving numerical stability if desired. (iii) We employ a memory-distributed storage of the environments needed for local updates. (iv) We accelerate local updates using randomized projections.  
Below, we briefly address each of these points.

(i) Our parallel MPO-MPO contraction follows a strategy first introduced by Stoudenmire and White 
\cite{PDMRG} in the context of MPO-MPS contractions needed for Density Matrix Renormalization Group (DMRG) ground state searches.
We partition the tensor train into multiple subsections (sub-TTs) and each node works serially on contracting two sub-TTs. During the computation, the nodes communicate via Message Passing Interface (MPI) in order to handle the bond shared by neighboring nodes. 

(ii) We describe two different implementations for reconnecting neighboring sub-TTs to each other, 
using two different MPO gauge choices.
The first implementation, following \cite{PDMRG}, utilizes an extra 2-leg
diagonal tensor containing the inverse of singular values on the boundaries between sub-TTs to ensure global consistence.
This yields  near-ideal parallelization speedup across all system sizes. The second implementation avoids the  inversion of singular values, thereby improving numerical stability. However,
it requires some additional calculations for merging sub-TTs to ensure global consistence. It therefore yields excellent parallelization speedup only for large system sizes requiring several sweeps before convergence.

(iii) We employ distributed-memory parallelism, as the memory consumption of the MPO-MPO contraction algorithm scales with \(\order{\mathcal{L}\chi^3}\), where \(\chi\) is the MPO bond dimension. For MPO-MPO contraction, \(\chi\) is typically large. In addition to the distributed memory parallelism, shared memory parallelism can be used in the form of parallel linear algebra algorithms.

(iv) We show how the above parallelization strategies can be used in combination with other, independent, strategies for accelerating contractions. We use two such strategies,
both involving randomized projections, to accelerate (iv.1) the singular value decompositions (SVDs) needed for local 2-site updates involved in the fit algorithm, and (iv.2) the 
contraction of 3-leg environment tensors with 4-leg MPO tensors. We briefly discuss these in turn.

(iv.1) Local updates typically require bond expansion and therefore often are implemented as two-site updates. For MPO-MPO contractions, local 2-site updates involve an SVD with cost $\mathcal{O}(\chi^3d^6)$. This cost can be reduced to
$\mathcal{O}(\chi^3d^4)$ by using the Controlled Bond Expansion (CBE) scheme recently introduced in the context of DMRG ground state sweeps \cite{CBE}. In this work, we implement a variant of CBE that employs Randomized SVD (RSVD) (an idea suggested in Ref.~\cite{McCulloch2024}), which likewise has costs $\mathcal{O}(\chi^3d^4)$ but  is simpler to implement. 

(iv.2) 
The contraction of a 3-leg environment tensor with two or three 4-leg MPO tensors, needed when updating environments, has a computational cost of $\mathcal{O}(\chi^4 d^3)$. We use randomized projections to reduce the dimension of the MPO bonds being contracted from $\chi$ to $k$ (usually we choose $k=\mathcal{O}(\chi^{1/2})$), thus reducing the computational cost of the contraction to $\mathcal{O}(\chi^3kd^3)$.
Similarly, we use randomized projections to reduce the cost for contracting together two environment tensors and 4 MPO tensors, needed for two-site updates, from $\mathcal{O}(\chi^4d^4)$ to 
$\mathcal{O}(\chi^3 k d^4)$.

Finally, we remark that the above strategies (i-iv) can, with minor adjustments, also be used for MPO-MPS contractions.
Similarly, (i-iii) can be used together with other schemes for accelerating MPO-MPO contractions,
such as the SRC \cite{SRC} and RSI \cite{RSI} 
schemes mentioned earlier, now applied to the contraction of sub-TTs; but that lies beyond the scope of this work.

This paper is structured as follows.
In Section \ref{sec:mpompo}, we review different contraction schemes for MPOs. In Section \ref{sec:random}, we present randomized projections and how to use them to accelerate the fit algorithm. In Section \ref{sec:pmpompo}, we present the parallelized versions of the fit algorithm. Section \ref{sec:results} illustrates the performance of our algorithms by showing numerical benchmark results for random tensor trains.

\section{MPO-MPO contraction}\label{sec:mpompo}

In this section we introduce some notational conventions and review various schemes for computing MPO-MPO contractions.

\begin{figure}[ht]
\centering
\includegraphics[width=\linewidth]{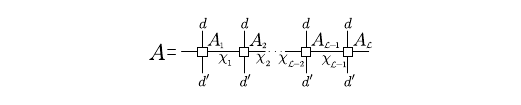}
\caption{Diagrammatic representation of the MPO 
$A = \prod_{\ell=1}^\mathcal{L} A_\ell$.}
\label{fig:mpo}
\end{figure}

An MPO of length $\mathcal{L}$ is a contraction $A = \prod_{\ell=1}^\mathcal{L} A_\ell$ of an ordered collection of degree-4 tensors $\{A_1,A_2,\dots,A_\mathcal{L}\}$, 
depicted schematically in \Fig{fig:mpo}.
The tensor elements are assumed to be real or complex numbers.
Each tensor $A_\ell$ has dimensions $(\chi_{\ell-1},d_{\ell},d'_{\ell},\chi_{\ell})$, where $d_{\ell}$ and $d'_{\ell} $ are the physical dimensions corresponding to the output and input indices, 
respectively, and $\chi_\ell$ is the dimension of bond $\ell$ connecting $A_\ell$ and $A_{\ell+1}$. Bonds represent contractions, i.e., summation over the shared indices.  

An MPO-MPO contraction $C = AB$ is defined as the contraction
of the output indices of $B$ with the corresponding input indices of $A$, as schematically depicted in \Fig{fig:mpo_mpo}(a). The goal of ``computing an MPO-MPO contraction'' is to find an MPO representation of $C$, depicted in \Fig{fig:mpo_mpo}(b). In other words, given the  local tensors $\{A_\ell\}$ and $\{ B_\ell \}$ representing $A$ and $B$, the goal is to find the local tensors $\{C_\ell\}$ representing $C$.
Typically, approximations are involved, such that $C \approx AB$.

To simplify the notation, we set \(\chi_0 = \chi_\mathcal{L} = 1\), and define the bond dimension of \(A\) as  \(\chi_A = \max_\ell\{\chi_\ell\}\).
When analyzing algorithms for finding approximations $C \approx AB$, we adopt the following simplifying conventions:
\begin{enumerate}
    \item We denote the largest bond dimension across all MPOs as $\chi=\max\{\chi_A,\chi_B,\chi_C\}$.
    \item We assume that all local physical dimensions are equal, i.e.\
    $d = d_{\ell}= d'_\ell$. 
    \item We assume the cost of a matrix multiplication between an $n_1\times n_2$ and an $n_2\times n_3$ matrix to be $\mathcal{O}\bigl(n_1n_2n_3\bigr)$.
\end{enumerate}

In computational practice, MPO contractions arise in various contexts. For example, consider a linear operator represented by a large matrix $A(x,y)$ of dimension $d^\mathcal{L}\times d^\mathcal{L}$, where $x=(x_1,\dots,x_\mathcal{L})$ and  $y=(y_1,\dots,y_\mathcal{L})$ are composite indices and each subindex $x_\ell, y_\ell$ ranges over ${1,\dots,d}$. This matrix can be represented as
an MPO $A = \prod_\ell A_\ell$ with output indices $x$ and input indices $y$. 
The contraction of two such MPOs, \(A\) and \(B\), then represents the matrix product
$\sum_y A(x,y) \, B(y,z)$.

MPOs can also represent multidimensional functions.
For example, when discretizing function variables
using the binary quantics representation, \cite{TT, ODlogN, Ritter2024}, one obtains MPOs with \(d = 2\).
For two-dimensional functions $f$, $g$, the contraction of their MPOs represents the integral $\int \md y \, f(x,y)\, g(y,z)$.

\subsection{Naive contraction}\label{ssec:naive}

A simple approach for obtaining an MPO representation for $C = AB$ is to contract the cores of $A$ and $B$ exactly site by site, $C_\ell = A_\ell B_\ell$, and then truncating the bond dimensions of the resulting $C_\ell$ tensors, if needed. This approach is straightforward in principle and exact; however, it is also naive, since in practice its computational costs can become prohibitive.

\begin{figure}[ht]
\centering
\includegraphics[width=\linewidth]{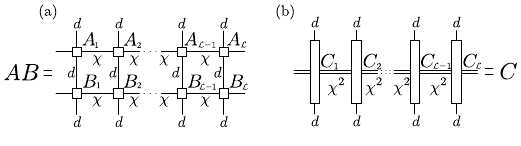}
\caption{Diagrammatic representations of (a) the contraction of two MPOs, $A= \prod_\ell A_\ell$ and $B = \prod_\ell B_\ell$, 
and (b) the MPO $C = \prod_\ell C_\ell$ obtained from an naive contraction. The double lines represent the 
bond dimensions of the $C_\ell$ tensors obtained from the  Kronecker product of $A_\ell$ and $B_\ell$.}
\label{fig:mpo_mpo}
\end{figure}

Concretely, $C_\ell = A_\ell B_\ell$
is the Kronecker product of $A_\ell$ and $B_\ell$,  depicted in \Fig{fig:mpo_mpo} (b)
and described in \Alg{alg:naive}.
Given $A_\ell$ and $B_\ell$ of size ($\chi_A\times d\times d\times \chi_A$) and ($\chi_B\times d\times d\times \chi_B$), we obtain $C_\ell$ as
\begin{align*}
    C_\ell(i_a, i_b, \sigma'', \sigma, j_a, j_b) = \sum_{\sigma'} A_\ell(i_a,\sigma'',\sigma',j_a) B_\ell(i_b,\sigma',\sigma,j_b),
\end{align*}
then reshape this as a ($\chi_A\chi_B \times d\times d\times \chi_A\chi_B$) tensor.
This results in bond dimensions growing multiplicatively, $\chi_C = \chi_A \cdot \chi_B$.

For algorithms involving numerous MPO-MPO contractions performed in succession, naive contractions
quickly become impractical, since the bond dimensions increase multiplicatively with each contraction.
Therefore, the output bonds have to be compressed (e.g.\ via truncated SVD) after each contraction to keep their size manageable. But even if the output has a low bond dimension, the algorithm remains expensive in terms of runtime and memory. 
Concretely, naive exact contractions have a computational complexity scaling
as $\mathcal{O}\bigl(\chi^4 d^3 \mathcal{L}\bigr)$, 
yielding MPO tensors with bond dimension $\chi^2$ and \(\mathcal{O}\bigl(\chi^4d^4)\) elements. 
For instance, with $d=2$ and $\chi=100$, the tensor that needs to be compressed requires about 12 GB of memory, making this approach unsuitable for large-scale problems. 
The SVDs required for compressing the MPO bond dimensions back to $\chi$ have an total computational cost that is even higher, $\mathcal{O}\bigl(\chi^6d^6\mathcal{L}\bigr)$. 
This limitation highlights the need for faster, less memory-intensive contraction schemes.

\begin{algorithm}
\caption{Naive Contraction (see Fig.~\ref{fig:mpo_mpo})} \label{alg:naive}
\begin{algorithmic}
\STATE \textbf{Input:} Tensor trains $A$ and $B$, tolerance $\tau$, maxbonddim $\chi_{\max}$
\FOR{$\ell$ in $1,\dots,\mathcal{L}$}
    \STATE $C_\ell \leftarrow$ contract $A_{\ell}$ and $B_{\ell}$ on their shared leg
\ENDFOR
\STATE $C \leftarrow$ compress($C$, $\tau$, $\chi_{\max}$)
\STATE \textbf{Output:} $C$
\end{algorithmic}
\end{algorithm}

\subsection{Zip-up contraction}\label{ssec:zipup}

The zip-up algorithm offers a fast, one-pass procedure for contracting two MPOs \cite{2sFit}. Its main advantage is efficiency, though it does not provide strict global error control.

\begin{figure}
\includegraphics[width=\linewidth]{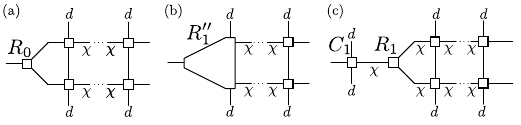}
\caption{Diagrammatic representation of the zip-up contraction.}
\label{fig:zipup}
\end{figure}

Given two MPOs $A$ and $B$, the algorithm 
for a left-to-right zip-up contraction proceeds as follows (see Fig.~\ref{fig:zipup} and Alg.~\ref{alg:zip-up}):
\begin{enumerate}
  \item \textbf{Initialization.} Start with a trivial tensor $R_0$ of shape $(1\times1\times1)$, with $R_0(1,1,1)=1$ (Fig.~\ref{fig:zipup}(a)). 
  \item \textbf{Local contraction.} For each site $\ell=1,\dots, \mathcal{L}$:
  Contract $R_{\ell-1}$ with the local tensors $A_\ell$ and $B_\ell$ to produce the intermediate 5-legged tensor $R_\ell''$ (Fig.~\ref{fig:zipup}(b)).
  \item \textbf{Compression.} Reshape $R_\ell''$ into a \((\chi d^2)\times(\chi^2)\) matrix grouping its indices: one side combines the left bond with the two physical legs, the other side combines the two right bonds. Apply a truncated SVD 
  to factorize this matrix. The left factor defines the new MPO tensor $C_\ell$, while the right factor becomes the updated residual $R_{\ell}$ (Fig.~\ref{fig:zipup}(c)).
  The bond dimension of the new $C_\ell$ can be chosen dynamically, if desired to achieve some target accuracy, by retaining all singular values larger than some specified truncation threshold.
\end{enumerate}

Before applying the zip-up contraction, the TTs should be brought into the right-canonical form, where every tensor except the site $\ell=1$ is right-orthogonal (see Fig.~\ref{fig:orth}(b)).
This gauge choice increases accuracy significantly without much additional cost.
Analogously, a right-to-left zip-up contraction can be performed by bringing the TTs into left-canonical form.

\begin{algorithm}
\caption{Zip-up Contraction (left-to-right)
(see Fig.~\ref{fig:zipup})
\label{alg:zip-up}}
\begin{algorithmic}
\STATE \textbf{Input:} Tensor trains $A$ and $B$, tolerance $\tau$, maxbonddim $\chi_{\max}$
\STATE $R_0 \leftarrow [[[1]]]$
\FOR{$\ell$ in $1,\dots,\mathcal{L}$}
    \STATE $R_\ell' \leftarrow$ contract $R_{\ell-1}$ and $A_{\ell}$ on their shared leg
    \STATE $R_\ell'' \leftarrow$ contract $R_\ell'$ and $B_{\ell}$ on their shared legs
    \STATE $M_\ell \leftarrow$ reshape $R_\ell''$ into matrix
    \STATE $U, S, V^\dagger \leftarrow$ SVD($M_\ell$, $\tau$, $\chi_{\max}$)
    \STATE $C_\ell \leftarrow U$ reshaped as tensor
    \STATE $R_\ell \leftarrow SV^\dagger$ reshaped as tensor
\ENDFOR
\STATE $C_\mathcal{L} \leftarrow C_\mathcal{L}SV^\dagger$
\STATE \textbf{Output:} $C$
\end{algorithmic}
\end{algorithm}

This procedure contracts the entire MPO in a single left-to-right sweep. The computational cost of computing $R_l''$ is $\mathcal{O}(\chi^4d^3)$ and the SVD costs $\mathcal{O}(\chi^4d^4)$. Therefore, the cost of the whole sweep is $\mathcal{O}\bigl(\chi^4d^4\mathcal{L}\bigr)$. This can be reduced to $\mathcal{O}\bigl(\chi^4d^3\mathcal{L}\bigr)$ using  randomized SVDs (Sec. \ref{ss:rsvd}).

\subsection{Fit contraction}\label{ssec:fit}

The fit algorithm approximates the product of two matrix product operators (MPOs) $A$ and $B$ by a new MPO $C$ solving the least-squares problem
\begin{equation}
    C = \underset{C}{\arg\min} \ \|A\, B - C\|_F^2, \label{eq:fit_min}
\end{equation}
where $\|\cdot\|_F$ denotes the Frobenius norm. Rather than solving Eq. (\ref{eq:fit_min}) globally, the algorithm proceeds by a sequence of local two-site optimizations. At each step two neighbouring sites $C_\ell,C_{\ell+1}$ are updated while all other tensors are kept the same \cite{1sFit}.
Since each such step incorporates information from the left and right environments of the tensors being updated, the fit algorithm is more accurate than the zip-up algorithm.
Though the details of the fit algorithm are well known, we briefly review its ingredients below, to set the stage for subsequent sections.

\subsubsection{Orthogonal tensors and canonical forms}\label{sss:canon}

\begin{figure}[ht]
\includegraphics[width=\linewidth]
{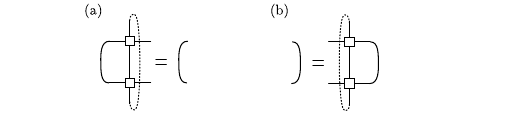}
\caption{Diagrammatic representation of the conditions for (a)  left orthogonality and (b) right orthogonality.}
\label{fig:orth}
\end{figure}

Orthogonal tensors play a crucial role in the fit algorithm. A tensor $O$ is said to be left orthogonal if $O^\dagger O = I$, with the contraction scheme shown in Fig.~\ref{fig:orth}(a). Analogously, it is said to be right orthogonal if $OO^\dagger = I$ with the contraction scheme shown in Fig.~\ref{fig:orth}(b).

A given tensor train may admit multiple equivalent representations that differ in the orthogonality properties of their sites. This is known as ``gauge freedom''. To standardize the representation, several canonical forms have been introduced, each imposing specific orthogonality conditions on the TT sites.
For convenience, we briefly recall some key properties of each below.

\paragraph*{Site-canonical form} 
A TT of the form $A = \prod_{\ell=1}^\mathcal{L} A_\ell$ is in site-canonical form w.r.t.\ site $\ell$ if it has an  orthogonality center (OC) at site $\ell$, meaning that every tensor $A_{\ell'}$ to the left or right of the OC ($\ell'< \ell$ or $\ell'> \ell$) is left- or right-orthogonal, respectively.
To obtain such a TT, one needs to perform successive QR decompositions from sites $1$ to $\ell-1$ and RQ decompositions from sites $\mathcal{L}$ to  $\ell+1$.

\begin{figure}
\includegraphics[width=\linewidth]{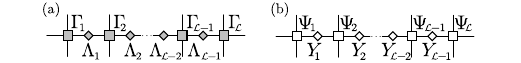}
\caption{Diagrammatic representation of an MPO in (a) the Vidal canonical form and (b) the inverse canonical form. 
}
\label{fig:vidal_inverse}
\end{figure}

\paragraph*{Vidal canonical form}
In this canonical form \cite{Vidal_2003}, one factors each 4-leg tensor into a 4-leg tensor $\Gamma$ and a diagonal singular-value matrix $\Lambda$. The Vidal decomposition uses $\Gamma$ and $\Lambda$ such that every $\Lambda_\ell \Gamma_{\ell+1}$ is left-orthogonal and $\Gamma_\ell \Lambda_\ell$ is right-orthogonal (Fig.~\ref{fig:vidal_inverse}(a)).
Therefore, every tensor combination $\Lambda_\ell\Gamma_\ell\Lambda_{\ell+1}$ constitutes an OC, as can be seen via the identifications $A_{\ell'} = 
\Lambda_{\ell'-1} \Gamma_{\ell'}$ for $\ell'< \ell$
and $A_{\ell'} = \Gamma_{\ell'} \Lambda_{\ell'}$
for $\ell'> \ell$.

\paragraph*{Inverse canonical form} 
This canonical form stores $Y_\ell=\Lambda_\ell^{-1}$ and defines the 4-leg
tensors $\Psi_\ell$ as $\Psi_\ell = \Lambda_{\ell-1}\Gamma_\ell\Lambda_\ell$.
Similarly to the Vidal form, every $\Psi_\ell Y_\ell$ is left orthogonal, every $Y_{\ell}\Psi_{\ell+1}$ is right orthogonal, and every site tensor $\Psi_\ell$ constitutes
an OC
(Fig.~\ref{fig:vidal_inverse}(b)). 
Below we denote the inverse canonical form components of $A$ by $\{\Psi_{A,\ell},Y_{A,\ell}\}$, etc. 
Because $\Lambda_\ell$ is diagonal, forming $Y_\ell$ and $\Psi_\ell$ is inexpensive, and before inverting $\Lambda_\ell$, very small singular values can be truncated for stability. 
However, the singular values have to be computed with high accuracy, using, e.g., a special recursive SVD algorithm (see App.~A of \cite{PDMRG}). Moreover, situations can arise where
the inversion of small singular values is unsafe \cite{Hastings}. Therefore, in Sec.~\ref{sec:pmpompo} below we will present two parallelization schemes, using either the site-canonical form or the inverse canonical form.

\subsubsection{Two-site local optimization}

\begin{figure}[ht]
\centering
\includegraphics[width=\linewidth]{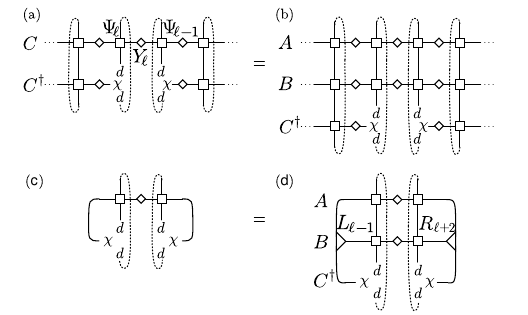}
\caption{ 
(a,b) Diagrammatic depictions of the left and right sides of \Eq{eq:fit_pde}, specifying a two-site update in the fit algorithm for contracting two MPOs in the inverse canonical form. (c,d) Compact forms of (a,b), obtained 
by contracting all tensors outside the two-site window associated with sites $\ell$ and $\ell+1$. (c) follows from (a) by exploiting the left orthogonality of $\Psi_{\ell'} Y_{\ell'}$ for $\ell'< \ell$  and 
the right orthogonality of $Y_{\ell''} \Psi_{\ell''+1} $ for $\ell''> \ell+1$. In (d), right- and left-pointing triangles represent the left and right environment tensors
$L_{\ell-1}$ and $R_{\ell+2}$, respectively (see Fig.~\ref{fig:envs}).
Dashed lines are equivalent to continuous one and are used to lighten the figure. 
}
\label{fig:derivatives}
\end{figure}

Expanding the cost function $\|AB-C\|_F^2$ gives
\begin{equation}
\| (AB) - C \|_F^2 = C^\dagger C - (AB)^\dagger C - C^\dagger (AB) + \text{const}. 
\end{equation}
Taking the derivative with respect to $(C_\ell C_{\ell+1})^*$ and setting it to 0 to find the minimum yields the linear system
\begin{equation}\label{eq:fit_pde}
\frac{\partial [C^\dagger C]}{\partial (C_\ell C_{\ell+1})^*} = \frac{\partial [C^\dagger (A B)]}{\partial (C_\ell C_{\ell+1})^*},
\end{equation}
which we solve exactly for each $C_\ell C_{\ell+1}$, iterating over $ \ell = 1, \dots, \mathcal{L}-1 $.
The left and right sides of Eq. (\ref{eq:fit_pde}) can both be expressed as contracted tensor networks with ``punched holes'' resulting from the removal of $C_\ell^*$ and $C_{\ell+1}^*$ \cite{XTRG}, as depicted in Figs.~\ref{fig:derivatives}(a) and \ref{fig:derivatives}(b), respectively.
Contracting together all tensors outside the two-site window  associated with sites $\ell$ and $\ell+1$ yields
Figs.~\ref{fig:derivatives}(c) and \ref{fig:derivatives}(d), respectively. For the former, 
we assumed that $C$ is in site-canonical form with OC on either $C_\ell$ or $C_{\ell+1}$, so that the orthogonality relations of \Fig{fig:orth} can be evoked recursively. For the latter, left and right environment tensor were introduced, which we discuss next.

\subsubsection{Environments} \label{sss:env}

\begin{figure}[ht]
\centering
\includegraphics[width=\linewidth]{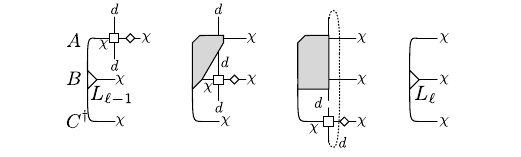}
\caption{Construction of the left environment in the inverse canonical form. 4-leg tensors depict $\Psi_{A,\ell},\Psi_{B,\ell},\Psi_{C,\ell}$, diamonds depict $Y_{A,\ell},Y_{B,\ell},Y_{C,\ell}$. In the site-canonical form, the 4-leg tensors represent $A_\ell,B_\ell,C_\ell$ and no diamonds occur. For the first, second, and third diagrams, the computational costs for contracting the four-leg tensor (square box) onto the rest of the diagram 
are $\mathcal{O}(\chi^4 d^2)$, $\mathcal{O}(\chi^4 d^3)$, and $\mathcal{O}(\chi^4 d^3)$, respectively.}
    \label{fig:envs}
\end{figure}

In  \Fig{fig:derivatives}(d), the left environment $L_{\ell-1}$ is the contraction of all tensors from site $1$ up to $\ell-1$; similarly $R_{\ell+2}$ is a contraction of the tensors on sites $\ell+2$ through $\mathcal{L}$.
Irrespective of the canonical form used, environments are composed of orthogonal tensors.
We start from  $L_0$ and $R_{\mathcal{L}+1}$, trivial 3-leg tensors of size $(1\times1\times1)$ with $L_0[1,1,1]=1$ and $R_{\mathcal{L}+1}[1,1,1]=1$. The environment tensors are then built sequentially by contracting one new site at a time (see Fig.~\ref{fig:envs}).
Once $L_{\ell-1}$ and $R_{\ell+2}$ are available, 
the 6-leg tensor of Fig.~\ref{fig:derivatives}(c) is obtained by contracting them with the local tensors 
from $A$ and $B$ on sites $\ell,\ell+1$.

Constructing an environment scales as $\mathcal{O}(\chi^4 d^3)$, and building the sequence of environments is one of the main contributions to the computational cost of the fit algorithm.

\subsubsection{Local update}

The locally optimized 
version of $C$ is obtained from the right side of Eq.~(\ref{eq:fit_pde}), involving the contraction
of four 4-leg MPO tensors and two environment tensors
to obtain the 6-leg tensor depicted in Fig.~\ref{fig:derivatives}(c), at a computational cost of $\mathcal{O}(\chi^4d^4)$.
We then reshape the 6-leg tensor as a $(\chi d^2)\times(\chi d^2)$ matrix and perform a truncated SVD on it to identify the two tensors \(C_\ell\) and \(C_{\ell+1}\) in site-canonical form, or the three tensors \(\Psi_{C, \ell} \), \(Y_{C, \ell}\) and \(\Psi_{C, \ell+1}\) in inverse canonical form.

A full SVD costs $\mathcal{O}(\chi^3d^6)$, whereas a randomized SVD would cost only $\mathcal{O}(\chi^3d^4)$ (see Sec. \ref{sec:random}).
We truncate the SVD either by keeping a number of singular values equal to the maximum bond dimension, or by discarding the singular values such that the error is below $\tau / \mathcal{L}$, where $\tau$ is a specified tolerance. (The division by $\mathcal{L}$ ensures
that the cumulative error from all $\mathcal{L}$ 
MPO tenors is comparable to the specified tolerance.) We estimate the error of truncating the SVD by the square root of the sum of the discarded singular values.

Having factorized the two-site tensor, we can identify the updated core tensors of \(C\): In site-canonical form, during a left-to-right (right-to-left) sweep, we update \(C_\ell = U\) (\(C_\ell = US\)) and \(C_{\ell+1} = SV^\dagger\) (\(C_{\ell+1} = V^\dagger\)), 
such that the canonical center containing \(S\) is always in the next pair of sites to be updated 
(Alg.~\ref{alg:site-canonical-update}).
In the inverse canonical form, we set $\Psi_{C,\ell} = US$,$\Psi_{C,\ell+1} = SV^\dagger$ and $Y_{C,\ell}=S^{-1}$
(Alg.~\ref{alg:inverse-canonical-update}).

\begin{algorithm}
\caption{ \quad 
2-site update in site-canonical form 
\label{alg:site-canonical-update}
}
\begin{algorithmic}
\STATE \textbf{Input:} $\ell$, TTs as $A$, $B$ and $C$ Envs $L$ and $R$, $\tau$, maxbonddim $\chi_{\max}$, direction
\STATE RHS$ \leftarrow$ contract $L$, $A_\ell$, $B_\ell$, $A_{\ell+1}$, $B_{\ell+1}$, $R$ on shared legs
\STATE $U, S, V^\dagger \leftarrow$ SVD(reshape(RHS, $\chi d^2\times\chi d^2$), $\tau$, $\chi_{\max}$)
\IF{direction == Forward}
\STATE $C_\ell \text{, \;} C_{\ell+1} \leftarrow U\text{, \;} S V^\dagger$ 
\ELSE
\STATE $C_\ell\text{, \;} C_{\ell+1} \leftarrow US\text{, \;} V^\dagger$ 
\ENDIF
\end{algorithmic}
\end{algorithm}

\begin{algorithm}
\caption{\quad
2-site update in inverse canonical form (Fig.~\ref{fig:derivatives})
\label{alg:inverse-canonical-update}}
\begin{algorithmic}
\STATE \textbf{Input:} $\ell$, TTs as $\Psi_A, Y_A, \Psi_B, Y_B, \Psi_C, Y_C$ Envs $L$ and $R$, $\tau$, maxbonddim $\chi_{\max}$, direction
\STATE RHS $\leftarrow$ contract $L,\Psi_{A,\ell}$,$Y_{A,\ell}$,$\Psi_{B,\ell}$,$Y_{B,\ell}$,$\Psi_{A,\ell+1}$,$\Psi_{B,\ell+1}, R$ on shared legs
\STATE $U, S, V^\dagger \leftarrow$ SVD(reshape($RHS$, $\chi d^2\times\chi d^2$), $\tau$, $\chi_{\max}$)
\STATE $\Psi_{C,\ell} \leftarrow US$
\STATE $\Psi_{C,\ell+1} \leftarrow S V^\dagger$
\STATE $Y_{C,\ell} \leftarrow S^{-1}$
\end{algorithmic}
\end{algorithm}

\subsubsection{Algorithm}

The variational fit algorithm (Alg.~\ref{a:fit}) requires an initial guess. This can be set as the result of the zip-up algorithm, or as one of the two TTs to contract. The algorithm starts by precomputing all right environments from right to left. Then, one sweeps left-to-right and right-to-left until convergence or a maximum number of sweeps is reached. At each step of a left-to-right sweep, one computes $1$ left environment and performs $1$ local update. Similarly, at each step of a right-to-left sweep, one computes $1$ right environment and performs $1$ local update. 

Compared to the zip-up algorithm, the fit algorithm achieves better error control and higher accuracy because the sweeping procedure incorporates more global information into each local update. 

\begin{algorithm}
\caption{Fit Contraction \label{a:fit}}
\begin{algorithmic}
\STATE \textbf{Input:} Tensor trains $A$ and $B$, $\tau$, maxbonddim $\chi_{\max}$
\STATE left\_orthogonalize(A)
\STATE left\_orthogonalize(B)
\FOR{$\ell$ in $\mathcal{L},\dots,3$}
    \STATE $R_\ell \leftarrow$ right\_environment($R_{\ell+1}, A_\ell, B_\ell, C_\ell$)
\ENDFOR
\STATE direction $\leftarrow$ Forward
\FOR{sweep in $1,\dots,N_{\text{sweeps}}$}
    \IF{direction $==$ Forward}
        \FOR{$\ell$ in $1,\dots,\mathcal{L}-1$}
            \STATE $L_{\ell-1} \leftarrow$ left\_environment($L_{\ell-2},A_{\ell-1},B_{\ell-1},C_{\ell-1}$) 
            \STATE update($\ell, A, B, C, L_{\ell-1}, R_{\ell+2}$, $\tau$, $\chi_{\max}$, direction)
        \ENDFOR
        \STATE direction $\leftarrow$ Backward
    \ELSE
        \FOR{$\ell$ in $\mathcal{L}-1,\dots,1$}
            \STATE $R_{\ell+2} \leftarrow$ right\_environment($R_{\ell+3},A_{\ell+2},B_{\ell+2},C_{\ell+2}$)
            \STATE update($\ell, A, B, C, L_{\ell-1}, R_{\ell+2}$, $\tau$, $\chi_{\max}$, direction)
        \ENDFOR
        \STATE direction $\leftarrow$ Forward
    \ENDIF
\ENDFOR
\STATE \textbf{Output:} $C$
\end{algorithmic} 
\end{algorithm}

\section{Randomized projection}\label{sec:random}

In this section we discuss randomized projections as a tool for reducing the computational costs of performing SVDs and contractions.
This tool is applicable when the tensors that are to be  projected exhibit approximate low-rank structure.

\begin{figure}[ht]
\centering
\includegraphics[width=\linewidth]{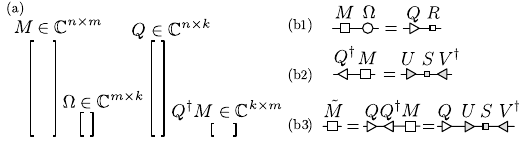}
\caption{(a) Visual representation of the sizes of the matrices involved in the randomized approximation.
(b) Graphical depiction of the equations 
used for randomized SVD:
(b1) $M\Omega = QR$, (b2) $Q^\dagger M = U S V^\dagger$, 
(b3) $\tilde M = Q Q^\dagger M = 
 \tilde U S V^\dagger $.
}
\label{fig:rsvd}
\end{figure}

The central idea is to insert randomized projection matrices on selected indices to compress the corresponding spaces. For example, given a matrix $M\in\mathbb{C}^{n\times m}$, we can approximate its column space by projecting it onto a random subspace spanned by the columns of $Q\in\mathbb{C}^{n\times k}$ with $k\ll \min(n,m)$, yielding a reduced matrix $Q^\dagger M\in\mathbb{C}^{k\times m}$ (see Fig.~\ref{fig:rsvd}(a)).
The projection matrix $Q$ is obtained by drawing a random test matrix $\Omega \in \mathbb{C}^{m\times k}$ (typically with independent and identically distributed Gaussian entries) and computing the orthonormal basis of $M\Omega$ via QR decomposition, $M \Omega = Q R$. 
The columns of $Q$ span a subspace that captures the dominant column space of $M$ with high probability \cite{RLA}, provided that the singular values of $M$ decay rapidly. By contrast, orthogonal tensors cannot be compressed this way, since all of their singular values are equal to one.

\subsection{Randomized singular value decomposition}\label{ss:rsvd}

Randomized projections can be used to perform a Randomized SVD (RSVD) \cite{RSVD}. Instead of performing an SVD on $M$, one 
approximates $M$ by the randomized projection, $\tilde M = Q Q^\dagger M$, and performs SVD on the reduced matrix $Q^\dagger M = U SV^\dagger$. 
 This leads to $\tilde M =  \tilde U S V^\dagger$, 
with $\tilde U = Q U \in \mathbbm{C}^{n \times k}$
(see Fig.~\ref{fig:rsvd}(b)), constituting an approximate SVD of $M$. Thus the computational cost is reduced from $\mathcal{O}(nm\min(n,m))$ 
for a full SVD to $\mathcal{O}(mnk)$, the complexity of evaluating $M\Omega$ and $Q^\dagger M$.

For standard truncated SVD, the approximation error in the Frobenius norm is the sum of the neglected singular values, i.e. $||M-QQ^\dagger M||_F = (\sum_{j>k}\sigma_j^2)^{1/2}$, where $\sigma_{j}$ are the discarded singular values of $M$. For randomized SVD, $(\sum_{j>k}\sigma_j^2)^{1/2}$ is only an lower bound for the error.
However, one may monitor the error and adapt $k$ to control the trade-off between cost and accuracy for any given problem.

\subsection{Randomized 2-site update and environments}\label{ss:rue}
\begin{figure}[ht]
\centering
\includegraphics[width=\linewidth]{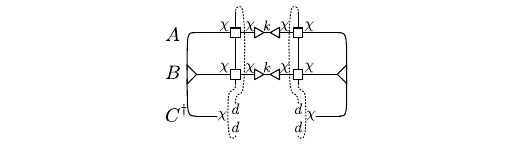}
\caption{Use of randomized projections to compute the 6-leg tensor representing the derivative of $C^\dagger (AB)$ with respect to $(C_\ell C_{\ell+1})*$, needed for a 2-site update. 
\label{fig:random_update}}
\end{figure}

In our tensor train setting, we can exploit randomized projections to accelerate the contractions of neighboring tensors, i.e., $A_\ell$ and $A_{\ell+1}$ and/or between $B_\ell$ and $B_{\ell+1}$.
To this end, we insert a separate randomized projector $Q Q^\dagger$ on each of the legs to be contracted  (see Fig.~\ref{fig:random_update}), 
then compute the corresponding reduced tensors, and perform the remaining contractions only at the end.
The randomized projection compresses the $A$-chain and/or $B$-chain bonds separately, accelerating the computation of the combined chain. Important caveat: this strategy is applicable only if each local tensor
whose bonds are being projected 
has a singular-value spectrum showing rapid decay also
above the tolerance spectrum.

\begin{figure}[ht]
\centering
\includegraphics[width=\linewidth]{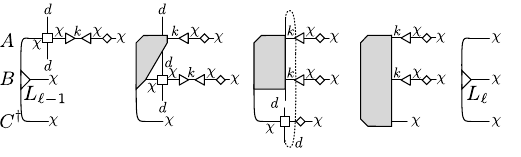}
\caption{Computation of environment tensor using two separate randomized projections
on its top and middle legs.
For the first, second, third, and fourth diagrams, the computational costs for performing
the contractions are $\mathcal{O}(\chi^3kd^2)$, $\mathcal{O}(\chi^2k^2d^2)$, and $\mathcal{O}(\chi^2k^2d^2)$, and $\mathcal{O}(\chi^3k)$, respectively.
\label{fig:random_env}}
\end{figure}

If either $A_\ell$ or $B_\ell$ can be approximated with a tensor of size $(\chi\times d\times d\times k)$, we can reduce the cost of contracting the 6-leg tensor 
of Fig.~\ref{fig:derivatives}(c) needed for
the 2-site update from $\mathcal{O}(\chi^4 d^4)$ to $\mathcal{O}(\chi^3kd^4)$ (see Fig.~\ref{fig:random_update}).
Similarly, the cost of building the environment at site $\ell$ can be reduced from $\mathcal{O}(\chi^4 d^3)$ to $\mathcal{O}(\chi^3kd^3)$ (see Fig.~\ref{fig:random_env}).
Usually, we set $k=\mathcal{O}(\chi^{1/2})$ because in our setting, the output MPO bond dimension is truncated to $\chi$. Therefore, it is reasonable to assume that each of the two incoming bonds contributes roughly a factor of $\chi^{1/2}$, leading to an overall effective rank of $\chi$ for the output.
A similar strategy has been implemented in \cite{zhangXTRG}, using truncated SVD instead of randomized projection.

Such a compression cannot be used to compute the environment in site-canonical form. The reason is that environments must be constructed from orthogonal tensors. Violating this orthogonality constraint leads to numerical instabilities. Therefore, we use randomized projection for environment compression only when employing the inverse canonical form.

Table \ref{t:cost_serial} shows the cost of the steps of the fit algorithm and its full cost. This breakdown of the cost highlights the benefit of random projection. In particular, we can see that two types of bottlenecks
can arise: 
(1) If $d$ is much smaller than $\chi$, the bottleneck lies in contractions of the environment and local update tensors, which both scale as $\mathcal{O}(\chi^4 d^3)$. 
Randomized projection reduces this cost to \(\mathcal{O}(\chi^3k d^3)\).
(2) If $d$ is comparable to or larger than $\chi$,  the bottleneck is the SVD factorization of the 6-leg two-site local update tensor, which scales as \(\mathcal{O}(\chi^3d^6)\).
Performing a randomized SVD instead, with $k=\chi$, decreases the cost to \(\mathcal{O}(\chi^3d^4)\). 

\begin{table}[ht]
    \centering
    \begin{tabular}{lcc}
         & without rand. & with rand.\\\midrule
    environment: & \(\mathcal{O}(\chi^4 d^3)\) & \(\mathcal{O}(\chi^3k d^3)\) \\
    local 2-site update:\\
    \hspace{1em}\textbullet~contraction & \(\mathcal{O}(\chi^4 d^4)\) & \(\mathcal{O}(\chi^3k d^4)\) \\
    \hspace{1em}\textbullet~SVD & \(\mathcal{O}(\chi^3 d^6)\) & \(\mathcal{O}(\chi^3 d^4)\) \\
    rand. proj. & $-$ & \(\mathcal{O}(\chi^3 d^2)\) \\
    total & \(\mathcal{O}(\chi^4d^4 + \chi^3 d^6)\) & \(\mathcal{O}(\chi^3kd^4)\)
    \end{tabular}
    \caption{
    Runtime complexity of computing the environment and performing a 2-site update step in the fit algorithm, with and without the use of randomized approximations, for a TT with physical dimensions $d$ and bond dimension $\chi$.}
    \label{t:cost_serial}
\end{table}
\section{Parallel fit contraction}\label{sec:pmpompo}

In this section, we discuss the parallelization of the fit algorithm. Among the existing MPO contraction schemes, it is the most naturally suited to parallel execution.

The naive contraction can be parallelized in a straightforward manner, but its scaling $\mathcal{O}(\chi^6)$ makes parallel speed-ups negligible. In contrast, the zip-up algorithm is inherently sequential: each step depends on all previous ones. We have explored a parallel zip-up scheme, where independent subsets of the tensor train are contracted and later merged. However, the intermediate bond dimensions grew far larger than the final solution, making the parallel version slower than a purely serial run.

We therefore focus here on two variants of the parallel fit algorithm: (I), a scheme based on the inverse canonical form, similar to those in Refs.~\cite{PDMRG,TDVP}; and (II), a new scheme in site-canonical form, avoiding explicit inversion of singular values.

In both cases, the central idea is to partition the TT into sub-TTs distributed across $N_{\text{procs}}$ processes, perform ``local'' sweeps (i.e.\ restricted to sub-TTs) in parallel, and reconcile boundary updates via MPI communication. This entails the following steps,
elaborated in more detail below: (1) Partition 
the TT to be constructed, $C$, into an even number of sub-TTs, labeled $n=1, \dots, N_{\text{procs}}$, distributed across $N_{\text{procs}}$ processes; (2) initialize the sub-TTs by moving the OCs of odd or even sub-TTs to their left- or right-most sites, respectively; (3) perform a set of local odd-forward/even-backward half-sweeps, during which all processes in parallel update their sub-TTs via a fit, sweeping forward for odd ones (moving their OCs to their right-most sites), and backward for even ones (moving their OCs to their left-most sites); (4) update the connections linking the OCs of neighboring odd-even sub-TTs; (5) perform a set of local odd-backward/even-forward half-sweeps, similar to the first, but with reversed sweeping directions;
(6) update the connections linking the OCs of neighboring even-odd sub-TTs; repeat steps (3-6) until convergence; and then (7) merge the final sub-TTs pairwise to obtain a fully merged final TT.

\subsection{Partitioning}

\begin{figure}
    \includegraphics[width=\linewidth]{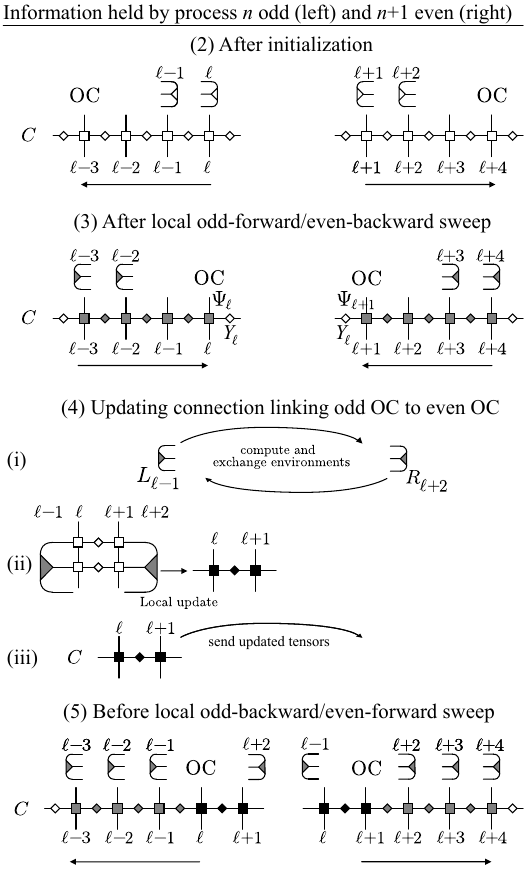}
    \caption{Illustrating steps (2) to (5) of the  parallel fit algorithm, for an odd-even pair of sub-TTs of a TT in inverse-canonical form. 
    (2) Showing the configuration after initialization: an odd sub-TT with OC at the left, an even sub-TT with OC at the right, and the  environments computed for each during initialization. (3) The final configuration after a local odd-forward/even-backward half-sweep: the OCs sit on the other sides of the sub-TTs; updated sites are shown with gray shading. (4) Updating the connection linking an odd OC to an even OC, involving the tensors
    $\Psi_\ell \Lambda_\ell \Psi_{\ell+1}$: (i) environments $L_{\ell-1}$ and $R_{\ell+2}$ are computed and exchanged; (ii) process  $n$ updates the tensors $\Psi_\ell \Lambda_\ell \Psi_{\ell+1}$;  (iii) updated tensors are sent to the neighboring process. (5) The initial configuration before the subsequent local odd-backward/even-forward half-sweep.
    }
    \label{fig:parallelfit}
\end{figure}

Partition the TT to be constructed for $C$ into
$N_\text{procs}$ sub-TTs, each containing some contiguous site and bond tensors.
Assign each sub-TT to a corresponding process, and the bond tensors linking two sub-TTs to both their processes [Fig.~\ref{fig:parallelfit}(a)].
If the input and output TTs had uniform bond dimensions across all bonds, one would divide the $\mathcal{L}$ site tensors as evenly as possible among $N_p$ processes. In practice, however, the bond dimensions are smaller near the left and right ends of the chain, where they are limited by the physical dimension. Therefore, we assign the left-most and right-most sub-TTs a few extra site tensors (typically 3 to 4) for load balancing. In the general case, a more refined strategy is possible by distributing site tensors according to the input and predicted output bond dimensions.

\subsection{Initialization}

Before the sweeps begin, we obtain an initial guess for the $C$ TT either through a zipup contraction or by setting the solution equal to one of the two input tensors.
Then, to prepare for the initial set of first local half-sweeps, each odd-numbered (say $n$) process moves the OC of its sub-TT to its left-most site and precomputes all right environments of that sub-TT, while each even-numbered process (say $n+1$) moves the OC of its sub-TT to its right-most site and precomputes all left environments on that sub-TT [Fig.~\ref{fig:parallelfit}(2)].
Since a full environment construction is as costly as a full sweep, we only build local environments from the tensors of each sub-TT. For a sub-TT starting at site $l_s$ and ending at site $l_e$, we initialize the boundary environments $L_{\ell_s-1}$ and $R_{\ell_e+1}$ as identities (or random orthogonal matrices). Building only the environment relative to their own sub-TT reduces the spatial cost from $\mathcal{O}(\chi^3\mathcal{L})$ to $\mathcal{O}(\chi^3\mathcal{L}/N_p)$. Though this degrades the accuracy in the first sweep, we find empirically that convergence is not hindered, and the solution quickly stabilizes in subsequent sweeps.

\subsection{Local odd-forward/even-backward half-sweeps}
\label{sweep:forward-backward}

Acting in parallel, all processes perform half-sweeps exactly as in the serial fit algorithm, but acting locally, each fitting only  its own sub-TT to the corresponding sub-TT representing $AB$. Each odd (say $n$) process sweeps forwards (left-to-right), moving its OCs to its right-most site, while each even (say $n+1$) process sweeps backwards (right-to-left), moving its OC to its left-most site [Fig.~\ref{fig:parallelfit}(3)].
After these sweeps, the OCs of the odd-$n$, even-$(n+1)$ pair of sub-TTs sit next to each other, say on sites $\ell$ and $\ell+1$. 
Moreover, process $n$ has updated all environments and bonds of its sub-TT up to and including environment $L_{\ell-2}$ and bond $\ell-1$; similarly, process $n+1$ has updated all environments and bonds of its sub-TT up to and including environment $R_{\ell+3}$ and bond $\ell+1$ [in Fig.~\ref{fig:parallelfit}(3), updated elements are shaded gray].

\subsection{Updating odd-OC to even-OC connections}
\label{connect:odd-even}

The connections linking the OCs of neighboring odd-even sub-TTs have not yet been updated. For the odd-$n$, even-$(n+1)$ pair of sub-TTs, 
the OCs sit on sites $\ell$ and $\ell+1$. The type of connection linking them depends on the choice of gauge. When using the site-canonical form, this connection is simply a bond; when using the inverse-canonical form, it is a bond-$Y_\ell$-bond [Fig.~\ref{fig:parallelfit}(3) shows the latter case]. To update these connections, the processes proceed as follows
[Fig.~\ref{fig:parallelfit}(4)]:

\begin{enumerate}[(i)]
    \item Process $n$ computes $L_{\ell-1}$ and sends it to process $n+1$; process $n+1$ computes $R_{\ell+2}$ and sends it to process $n$. This information is needed both for the next step (ii), and
    as initial data for the subsequent local odd-backward/even-forward half-sweeps \ref{sweep:backward-forward}.
    \item One of the two processes performs the two-site update on the sites $\ell$ and $\ell+1$, using the environments $L_{l-1}$ and $R_{l+2}$.
    When using site-canonical form this yields updated site tensors $C_l$, $C_{l+1}$; when using inverse-canonical form, this yields updated site tensors $\Psi_\ell$, $\Psi_{\ell+1}$ and an updated bond tensor $Y_\ell$. 
    \item The process with the updated tensors $C_\ell C_{\ell+1}$ 
    or $\Psi_\ell Y_\ell \Psi_{\ell+1}$ communicates them to its partner.
    
\end{enumerate}

\subsection{Local odd-backward/even-forward half-sweeps}
\label{sweep:backward-forward} 
All processes  perform another fit of their sub-TTs for $C$ to the corresponding sub-TTs for $AB$, but with reversed sweeping directions: odd processes sweep backwards, moving their OCs to their left-most sites, while even processes sweep forwards, moving their OCs to their right-most sites [Fig.~\ref{fig:parallelfit}(5)]. 

\subsection{Updating even-odd OC connections}
\label{connect:even-odd}
Then, the connections between even-odd OCs are updated, e.g.\
connecting even-$(n-1)$ to odd-$n$, or even-$(n+1)$ to odd-$(n+2)$. 

Steps \ref{sweep:forward-backward} to \ref{connect:even-odd} constitute a set of local full-sweeps. They are repeated until convergence with a specified tolerance or bond dimension has been achieved, say $N_s$ times.

\subsection{Merging}

After $N_s$ sets of local full-sweeps, the final TT is reconstructed. In the inverse canonical form, the local updates are gauge preserving. This means that if we merge the locally updated sub-TTs, we obtain the globally consistent TT.
Therefore, no further manipulations are required.
On the other hand, in site-canonical form, merging requires care: each sub-TT has its own canonical center, and the gauge of neighboring sub-TTs is not consistent. To merge two sub-TTs consistently, one needs to update the bond between the sub-TTs. The whole merge is done by pairwise merging of sub-TTs, requiring a $\mathcal{O}\left(\log(N_p)\right)$ number of merges.

Assume, for notational simplicity, that each sub-TT is of length $\Lsub = \mathcal{L}/N_p$. In the first merging round, we merge in parallel each sub-TT with neighboring sites $k_1^i\Lsub$ and $k_1^i\Lsub + 1$, with $k_1^i=1+2^1i, i=0,\dots,N_p/2-1$.
Merging across those two sites requires the environments $L_{k^i_1\Lsub -1}$ and 
$R_{k^i_1\Lsub + 2}$.
Then, in the second round, we merge in parallel across sites $k_2^i\Lsub$ and $k_2^i\Lsub + 1$, with $k_2^i=2+2^2i, i=0,\dots,N_p/4-1$. That requires $L_{k_2^i\Lsub-1}$, computed by starting from $L_{k_1^i\Lsub-1}$, and $R_{k_2^i\Lsub+2}$, computed by starting from $R_{k_1^{i+1} \Lsub+2}$.
We iterate these steps until we merge the last two remaining sub-TTs across the sites $\mathcal{L}/2$ and $\mathcal{L}/2 + 1$, requiring the environments $L_{\mathcal{L}/2-1}$ and $L_{\mathcal{L}/2+2}$. Given the sequential nature of the environments computation, the construction of those scales linearly in the total TT length $\mathcal{L}$, rather than in the local sub-TT size.

This overhead is negligible for contractions requiring many sweeps, but may hinder the scalability if only a small number of sweeps are performed.

\begin{algorithm}[ht]
\caption{Parallel Fit Contraction (see Fig.~\ref{fig:parallelfit}) \label{alg:parallelfit}}
\begin{algorithmic}
\STATE \textbf{Input:} TTs $A$ and $B$, initial guess $C$, tolerance $\tau$, maxbonddim $\chi_{\max}$, MPIrank, partition.
\STATE \textbf{Initialization:}
\IF{MPIrank mod $2 == 1$}
    \STATE direction $\leftarrow$ Forward
    \STATE move OC of sub-TT to its left-most site
    \STATE $R \leftarrow$ compute\_right\_environments($A, B, C$)
\ELSE
    \STATE direction $\leftarrow$ Backward
    \STATE move OC of sub-TT to its right-most site
    \STATE $L \leftarrow$ compute\_left\_environments($A, B, C$)
\ENDIF
\FOR{sweep in $1,\dots,N_{\text{sweeps}}$}
    \STATE \textbf{Local sweep:}
    \STATE sweep\_sub-TT() \COMMENT{As serial sweep, see Alg. \ref{a:fit}}
    \STATE \textbf{Update connection}
    \IF{direction $==$ Forward}
        \STATE $\ell \leftarrow$ last\_in(partition)
        \STATE $L_{\ell-1} \leftarrow$ compute\_left\_environment($L_{\ell-2}, A, B, C$) 
        \STATE send($L_{\ell-1}$); receive($R_{\ell+2}$)
        \STATE update($\ell, A, B, C, L_{\ell-1}, R_{\ell+2}$, $\tau$, $\chi_{\max}$, direction)
        \STATE send($\Psi_{C,\ell},Y_{C,\ell},\Psi_{C,\ell+1}$)
        \STATE direction $\leftarrow$ Backward
    \ELSE
        \STATE $\ell \leftarrow$ first\_in(partition) - 1
        \STATE $R_{\ell+2} \leftarrow$ compute\_right\_environment($R_{\ell+3}, A, B, C$)
        \STATE send($R_{\ell+2}$); receive($L_{\ell-1}$)
        \STATE receive($\Psi_{C,\ell},Y_{C,\ell},\Psi_{C,\ell+1}$)
        \STATE direction $\leftarrow$ Forward
    \ENDIF
\ENDFOR
\STATE \textbf{Output:} $C$
\end{algorithmic}
\end{algorithm}

\subsection{Theoretical scaling of the parallel algorithm}

The serial fit algorithm precomputes $\mathcal{L}$ environments, and then in each sweep it computes $\mathcal{L}\,N_s$ environments, and performs $\mathcal{L}\,N_s$ update.
In the parallel version, each process precomputes $\frac{\mathcal{L}}{N_p}$ environments and then in each sweep it computes $\bigl(\frac{\mathcal{L}}{N_p}-1\bigr)N_s$ environments, and performs $\bigl(\frac{\mathcal{L}}{N_p}-1\bigr)N_s$ updates. In the communication phase, ignoring the waiting time, $1$ environment is computed per process and $1$ update is performed.

The theoretical speedup is therefore ideal (i.e. using $N_p$ processes leads to a $N_p$ speedup).
However, for the site-canonical form, we need to take into consideration the merging step at the end of the sweeps. In the binary merge, an additional number of updates and environments needs to be computed. This leads to an added cost of $\mathcal{O}(\mathcal{L})$ environments and $\mathcal{O}\left(\log(N_p)\right)$ updates.

\section{Results} \label{sec:results}

In this section, we evaluate the speedup of the serial fit algorithm with and without randomized projections. We also evaluate the performance of the parallel fit contraction scheme by measuring the speedup achieved compared to the serial implementation. Speedup is the standard metric for evaluating the efficiency of a parallel algorithm, defined as the ratio between the runtime with a single process and the runtime with $N_{p}$ processes.

Both algorithms were implemented based on the tensor4all julia libraries \cite{tensor4all}. The parallel experiments were conducted on the LMU Physics computing cluster, on machines equipped with Intel CPUs with 6 cores, and 23 GB of RAM each. To maximize storage and minimize queueing time, the randomized fit experiments were run on the cluster partition of the LMU Physics computing cluster while requesting only 1 core, but allocating up to 1 TB of memory.

\subsection{Experimental setup}

In our experiments, tensor trains \(A\) and \(B\) were initialized using tensors with real numbers randomly sampled from a uniform distribution between $[\alpha,1]$, and then normalized. The parameter $\alpha \in [-1, 1]$ regulates the difficulty of the compression \cite{SRC}. Values of $\alpha$ close to $1$ or $-1$ yield TTs that are easier or more difficult to contract, respectively.

For the speedup analysis we utilized TTs of length 102 with bond dimensions ranging from 100 to 150 and an $\alpha=-0.5$.

For accuracy analysis, we needed to calculate the ``exact'' solution, which is computationally demanding. Therefore, these experiments were run with simpler TTs of length 102, with bond dimension $\chi=30$ and $\alpha=-0.33$.
These experiments were run for different numbers of sweeps $N_s$.

For the random projection time analysis, the experiments needed much larger bond dimensions to show the computational scaling, therefore, these were run with shorter TTs of 50 sites.

All of the experiments were run on 5 instances of randomly generated TTs, then the accuracy and speedup were plotted by averaging the 5 experiments.

\subsection{Numerical results of the parallel algorithm}

\begin{figure}[ht]
\includegraphics[width=\linewidth]{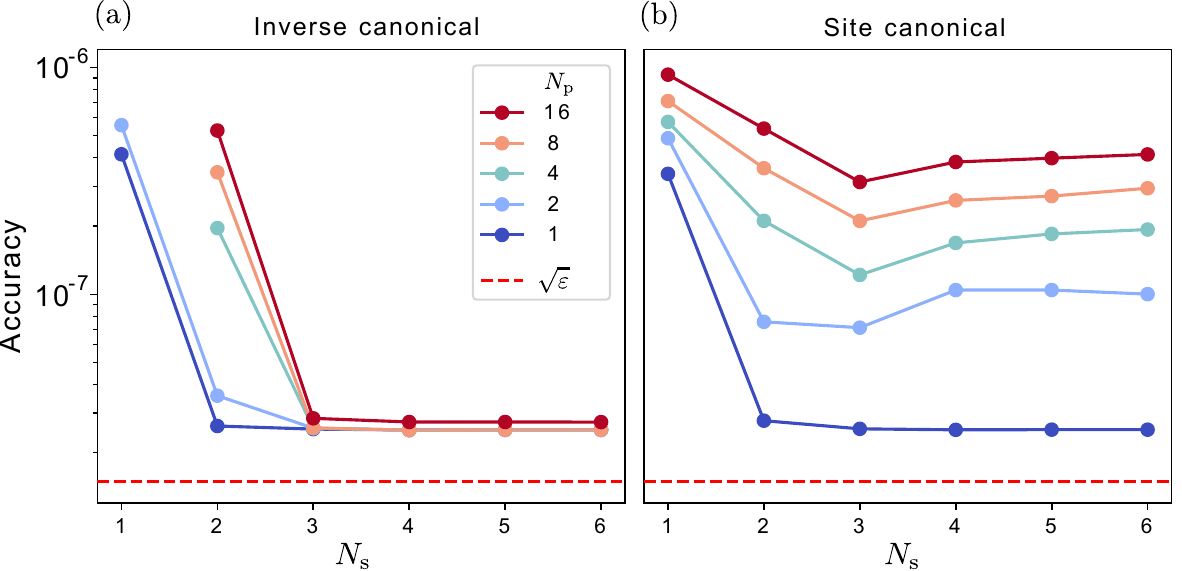}
\caption{Accuracy (averaged out of 5 runs with different seeds) plotted with respect to the number of sweeps $N_s$ using the inverse canonical and site-canonical form. The result obtained by using one or multiple processes is compared to the one obtained by using the naive contraction without truncation. The dashed red line is the square root of $\varepsilon$-machine, the highest accuracy achievable using the Frobenius norm. The Tensor Train has $\mathcal{L}=102$ and maximum bond dimension $\chi=30$. The tolerance was set to 0 to ensure the algorithm was as accurate as possible given a maximum bond constraint.}
\label{fig:accuracy}
\end{figure}

To evaluate the accuracy of the parallel approach, Fig.~\ref{fig:accuracy} shows the accuracy obtained by the serial and the parallel algorithms as a function of the number of sweeps.
The error is measured as the Frobenius norm of the difference between the tested TT and the one obtained using the naive contraction. The Frobenius norm of a TT is given by the square root of the TT contracted with its adjoint. Therefore, we compare the measured error with  $\sqrt{\varepsilon}$, where $\varepsilon$ is the machine precision.

\subsubsection{Accuracy of inverse canonical form}
Both in the serial and in the parallel version, the error decreases monotonically with the number of sweeps. Above 2 nodes, the error obtained after the first sweep is very big, therefore, those were left out of the plot. We can see that the serial and the parallel versions with 2 nodes have almost the same accuracy, converging within two steps, whereas using more nodes requires one extra step for convergence.

\subsubsection{Accuracy of site-canonical form}

The serial version converges after the second sweep, whereas the parallel versions show somewhat worse accuracy. We suspect this is given by the applications of multiple QR decompositions on the solution TT, operations that are not performed in the inverse canonical form.  We do not expect the accuracy to improve for $N_s>6$. If this were the case, the accuracy of the parallel version with 2 nodes would have converged before the sixth sweep.

\subsubsection{Speedup of inverse canonical form}
Figure~\ref{fig:su_all}(a) and \ref{fig:su_all}(b) report the speedup in inverse canonical form as a function of $N_p$ for different values of the maximum bond dimension $\chi$ and different amounts of sweeps. We can see that the scalability is quasi-ideal for any level of the bond dimension and for any number of sweeps. With larger bond dimensions and more sweeps, we see slightly better efficiency due to the increased computation-to-communication ratio.

\subsubsection{Speedup of site-canonical form}
\begin{figure}[ht]
\includegraphics[width=\linewidth]{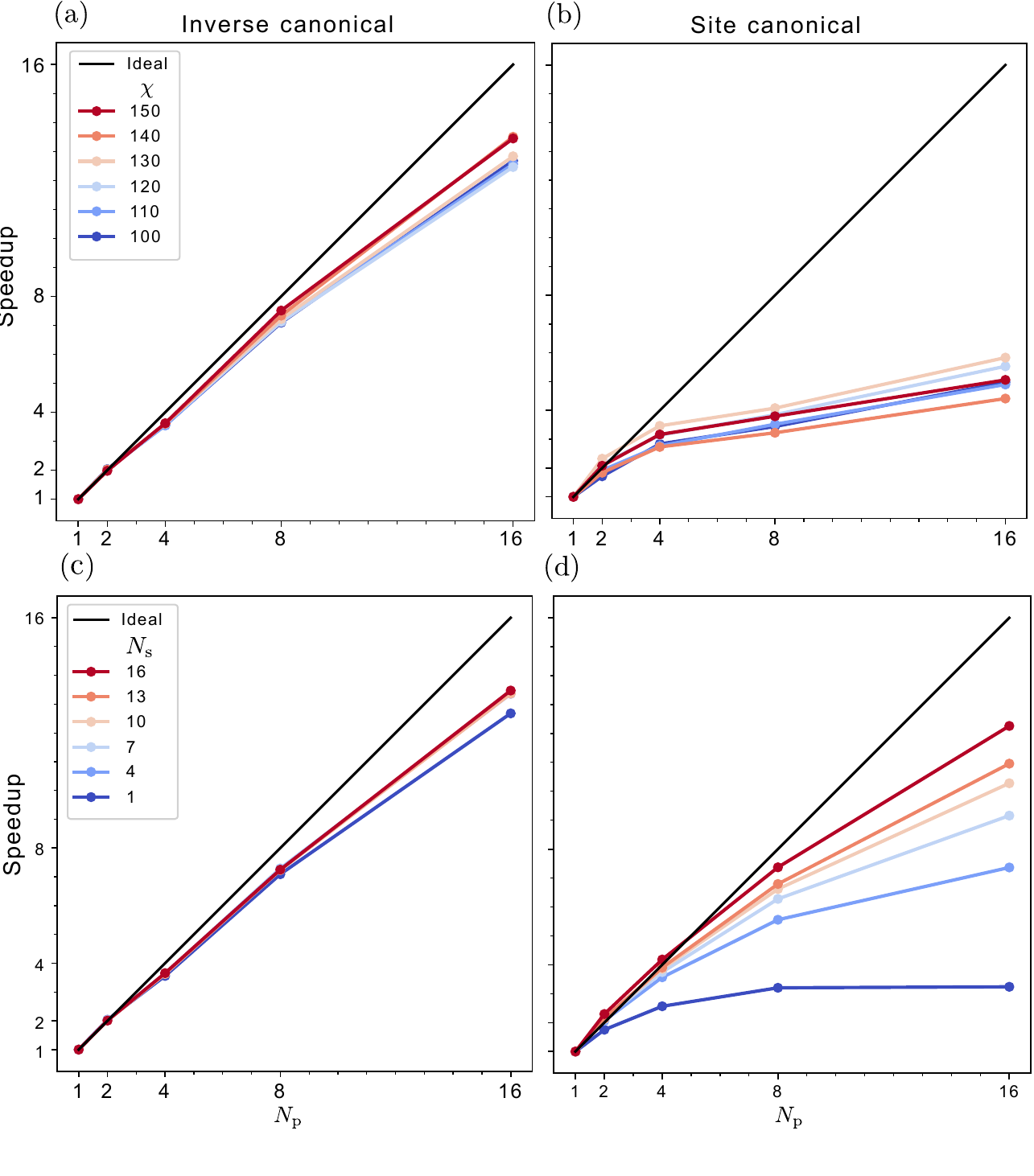}
\caption{Speedup (averaged out of 5 runs with different seeds) obtained by the parallel fit algorithm. The top-left (a) and top-right (b) plots show runs for different amounts of $\chi$, with a TT with $\mathcal{L}=102$ and $N_s=1$. The bottom-left (c) and bottom-right (d) plots show runs for different amounts of sweeps $N_s$, with a TT with $\mathcal{L}=102$ and $\chi=100$.}
\label{fig:su_all}
\end{figure}

Figures~\ref{fig:su_all}(c) and \ref{fig:su_all}(d) report the speedup in site-canonical form as a function of $N_p$ for different numbers of sweeps and different bond dimension $\chi$. Similar to before, we do not notice any major difference of the speedup with respect to the bond dimension. On the other hand, the number of sweeps heavily influences the speedup of the algorithm.
Although the parallelization achieves a 12x speedup with 16 processes when the algorithm runs for $16$ sweeps, the parallelization suffers when the scheme is run only for a low number of sweeps. Therefore, if one needs just a few sweeps, using a limited number of processes is recommended.

\subsubsection{Overall analysis}

When inverting the singular values is not problematic for the application, the inverse canonical form is the preferred choice: it maintains high accuracy, has a convergence comparable with the serial one, and exhibits quasi-ideal speedup even at large processes counts.
On the other hand, if inverting the singular values leads to error or numerical instability, then the site-canonical form provides an alternative. Its accuracy is slightly lower, but for a few nodes, it is still a viable option. Furthermore, if many sweeps are required and the error tolerance is fairly large, the parallel version provides a good speedup, making even larger node counts beneficial.

\subsection{Numerical results of the randomized fit}

\begin{figure}[ht]
\includegraphics[width=1\linewidth]{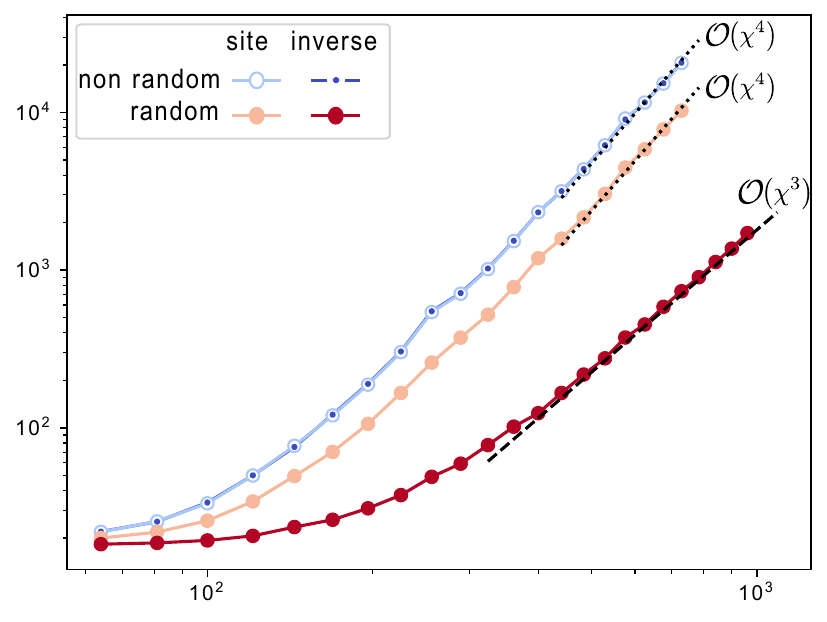}
\caption{Logarithmic plot of the runtime of fit algorithm on a of TTs with $\mathcal{L}=50$, run for 1 sweep using randomized and non-randomized schemes. For the site-canonical form, the randomized scheme involves RSVD and randomized 2-site updates (Fig.~\ref{fig:random_update}), for the inverse canonical form, the randomized scheme additionally involves projections when computing the environments (Fig.~\ref{fig:random_env}). For high bond dimensions, we can see the $\mathcal{O}(\chi^4)$ scaling when the environment is calculated exactly, whereas the leading computational cost is $\mathcal{O}(\chi^3)$ when the environment is calculated with random projections.}
\label{fig:randoms}
\end{figure}

Figure~\ref{fig:randoms} shows the comparison between non-randomized and randomized fit algorithms, in both canonical forms. The dashed and dotted black lines represent reference scaling $\mathcal{O}(\chi^3)$ and $\mathcal{O}(\chi^4)$ respectively.

In the environments computation, the cost of contracting the tensors scales as $\mathcal{O}(\chi^4)$. However, for small $\chi$, the leading cost is not the matrix-matrix multiplications per se, but the memory allocation needed to contract the tensors (see Fig.~\ref{fig:envs}), which scales as $\mathcal{O}(\chi^3)$.
For small $\chi$, the runtime curves start flat, and then for $\chi$ in the hundreds, they all show a $\mathcal{O}(\chi^3)$ scaling. Afterwards, for $\chi\approx500$, the algorithms that build the environments without randomized projections start scaling as $\mathcal{O}(\chi^4)$.

The non-randomized site canonical and inverse canonical fit take qualitatively the same time to compute one sweep, and they both scale as $\mathcal{O}(\chi^4)$ for large $\chi$. 
As already mentioned in Section \ref{ss:rue}, since the tensors needed to form the environments are orthogonal, one cannot use the randomized projection to speed up the computation of the environments in the randomized site-canonical fit.
Therefore, only the SVD and the 2-site update have been sped up, by adding projections as shown in Fig.~\ref{fig:random_update}. Fig.~\ref{fig:randoms} shows that the scaling is still $\mathcal{O}(\chi^4)$ for large $\chi$. Nonetheless, it is faster than the non-randomized counterpart.
In the randomized fit in inverse canonical form, assuming that the tensors needed to form the environments can be approximated as explained in Section \ref{ss:rue}, one can speed up the environments with randomized projection, by adding projections as shown in Fig.~\ref{fig:random_env}.
This makes the computational cost of the fit algorithm $\mathcal{O}(\chi^3k)$ also for large $\chi$. In the experiments, $k$ has been set as $\mathcal{O}(\chi^{1/2})$, however, the tensor sizes have been reduced so much that the leading cost is the memory allocation. We can see in the bottom right curve that randomized fit in inverse canonical form is notably faster compared to the other versions.

The accuracy of this method is heavily dependent on the problem. The error achieved is proportional to the singular values discarded during the randomized projection.
In the contraction of randomly generated TT, applying this approximation leads to an error of the same order of magnitude of the norm of the tensors themselves, since the singular values of the site tensors decay slowly.
However, if the singular values decay exponentially, as in \cite{zhangXTRG}, the error can reach orders of magnitude comparable to that of the non-randomized fit.
Nonetheless, the error of the randomized variant is proportional to the singular values discarded during the projection step.
Therefore, a practical strategy is to perform randomized sweeps initially, followed by non-randomized sweeps to ensure convergence to the desired solution.

\section{Summary and Outlook}
\label{sec:SummaryOutlook}

In this work, we have shown that the parallelization scheme based on the inverse canonical form introduced by Stoudenmire and White in the context of DMRG \cite{PDMRG} extends naturally and efficiently to MPO-MPO contractions. In this formulation, the parallel algorithm maintains an accuracy comparable to the serial implementation and exhibits quasi-ideal speedup across a wide range of bond dimensions and numbers of sweeps. When the inversion of singular values is numerically stable, this strategy is the preferred choice.

However, when the inversion of singular values leads to numerical instabilities, the site-canonical form provides a viable alternative, in particular, it delivers meaningful speedups when many sweeps are required.

Furthermore, we have also explored the integration of randomized linear algebra techniques into the fit algorithm. Our results show that randomized projections can significantly reduce the computational cost when the tensor sites admit low-rank approximations. While the effectiveness of this approach is problem-dependent, it provides further acceleration in applications with favorable spectral properties.

For future research, it would be beneficial to conduct a systematic comparison with the Successive Randomized Contraction (SRC) algorithm \cite{SRC}.
In particular, it is important to evaluate how the SRC scheme compare to the randomized fit variant presented in this work and whether they can be efficiently parallelized.

\section*{Acknowledgments}

The Flatiron Institute is a division of the Simons Foundation. 
This work was funded in part by the Deutsche Forschungsgemeinschaft under Germany’s Excellence Strategy EXC-2111 (Project No. 390814868). It is part of the Munich Quantum Valley, supported by the Bavarian state government with funds from the Hightech Agenda Bayern Plus.

\FloatBarrier
\bibliographystyle{elsarticle-num}
\bibliography{bib/bibliography}

\end{document}